\documentclass[pre,aps,twocolumn]{revtex4-1}
\usepackage{amsmath,natbib,amsfonts}
\usepackage{amssymb,color}
\usepackage{graphicx}
\newcommand{\beqa}{\begin{eqnarray}}
\newcommand{\eeqa}{\end{eqnarray}}

\begin{document}
\title{Single Particle Brownian Heat Engine With Microadiabaticity} 
\author{Arnab Saha$^1$, A. M. Jayannavar$^{2,3}$} 
\email{sahaarn@gmail.com, jayan@iopb.res.in}
\affiliation{$^1$Department of Physics, Savitribai Phule Pune University, Ganeshkhind, Pune  411007, India.\\
$^2$Institute of Physics, Sachivalaya Marg, Bhubaneshwar 751005, Odhisha, India.\\
$^3$Homi Bhabha National Institute, Training School Complex, Anushakti Nagar, Mumbai 400085, India. }

\date{\today}
\begin{abstract}
{\textcolor{black}{Micro-to-nano scale thermal devices that operate under large thermal fluctuations, are an active field of research where instead the average values, the full distributions of thermodynamic quantities are important. Here we study a model of stochastic heat engine consisting of a harmonically trapped Brownian particle driven by the time-periodic strength of the confinement, within two thermal baths of different temperatures. The particle follows two isotherms correspond to two baths and connected by two micro-adiabates. The microadiabaticity is implemented by conserving the phase space volume of the particle along the adiabatic paths. Here we show that it can operate as an engine or as a heater under microadiabaticity, depending on the parameter space. We also compute the distribution of stochastic efficiency and its averages for different cycle times of the engine.\\
{\bf{Keywords}}: Stochastic heat engine, Microadiabaticity. \\
{\bf{PACS}}: 5.70.Ln, 05.40.-a, 05.20.-y}}
\end{abstract}

\maketitle

\section{Introduction}
{\textcolor{black}{Thermodynamic heat engines of macroscopic world consume heat from a hot reservoir, convert it partially to useful work and deliver the rest to a cold reservoir, before the cycle repeats. At quasi static i.e. zero power limit, the efficiency of the engines are bounded by Carnot efficiency whereas at maximum power, the limit is given by Curzon and Ahlborn. The average thermodynamic quantities are enough to determine these efficiencies \cite{Callen}.}}

{\textcolor{black}{Current technology allow us to track and control the motion of a micro-to-nanometer sized particle in a solvent under external fields. Thermodynamics of such small systems is under intense scrutiny now \cite{broek2006,nakagawa,marathe2007,benjamin2008,seifert2008,engel2013,tu2014,tu2014,holubec2014,puglisi2015,rana2014,rana2016}. Typical energy scale of such systems is $\sim$ $K_BT$, where T is the ambient temperature and $K_B$ is the Boltzmann constant. Due to dominant thermal fluctuations, the thermodynamic features are different from that of large systems. Macroscopic thermodynamics is often inadequate to describe the thermodynamic behavior of small systems. It is now being analyzed with recently developed, novel thermodynamic principles (namely, stochastic thermodynamics \cite{udo1,sekimoto1998,seifert2005,dan2005,saikia2007,jop2008,saha2009,sekimoto2010,mahato2011}). It systematically incorporates the effects of fluctuations. In small scales, though the exchange of energy between the particle and its surroundings becomes stochastic, yet under the realm of stochastic thermodynamics, one can reformulate the concept of work, heat and entropy production for a given microscopic trajectory of the particle. Averaging these stochastic thermodynamic variables over all possible trajectories, we obtain macro-thermodynamics, together with a class of equalities, called fluctuation theorems \cite{udo1,ritort2003,harris2007,jarzynski2010,lahiri2014}, which are valid even far from equilibrium conditions.}}

{\textcolor{black}{Heat engines in macro-thermodynamics play a major role in applications and in conceptual development of the subject itself. In microscopic world, stochastic heat engines \cite{edgar1} are one of the major play grounds of stochastic thermodynamics. The cycle of the single particle engine of our concern here, consists of two isotherms of different of temperatures $T_h$ and $T_c$ ($T_h > T_c$), connected by two adiabates. Studies on Similar model systems where isotherms are connected by instantaneous adiabatic jumps \cite{rana1,rana2}, have revealed the fundamental differences between the thermodynamic features of macro- and micro- heat engines. Here we implement smooth, experimentally realizable, micro- adiabatic connections between the isotherms numerically. The adiabaticity is maintained by conserving the phase space volume of the Brownian particle, that ensures average heat dissipation to be zero over large cycle times \cite{martinez,ignacio}. Here we will focus on thermodynamic properties of Brownian heat engines driven by micro-adiabatic protocol.}}\\

\section{Model} 

We consider a one dimensional Brownian particle with position x and velocity v in under-damped condition, trapped within a harmonic potential with time dependent strength k(t) being the protocol for the thermodynamic force that acts on the particle. The equation of motion is given by
\beqa
m\frac{dv}{dt}=-\gamma v-k(t)x+\sqrt{\gamma T}\xi(t)
\eeqa
in units of $K_B$. We consider the mass m and friction coefficient $\gamma$ of the particle to be unity. $\xi$ is the delta correlated Gaussian white noise, mimicking the thermal environment due to the solvent surrounding the particle.

\section{Microadiabaticity} 
 
{\textcolor{black}{Here the single particle stochastic machine follows two isotherms correspond to two thermal baths and two adiabates, connecting them. Along the isotherms, the heat is exchanged between the particle and the corresponding thermal bath. Along the adiabatic steps, the protocol will be constructed in such a way that in quasistatic regime the average heat exchange between the particle and the bath will be zero, implying that the phase space volume and consequently the average entropy of the particle will be conserved along the adiabatic paths. In quasistatic regime, the particle is very close to equilibrium and we write the canonical partition function as $Z_t=\int dxdv\exp(-\beta_tH_t)$ where $\beta_t=\frac{1}{T_t}$ and $H_t=\frac{1}{2}mv^2+k_tx^2$. The suffix $t$ denotes that the quantity is changing slowly with time as a parameter. The free energy is evaluated as $F=-T_t\ln Z_t=-\frac{T_t}{2}\ln(\frac{4\pi^2T^2}{mk_t}).$ Along the adiabatic step, if we demand the change of entropy $\Delta S=-(\frac{\partial F_t}{\partial T_t})$ to be constant, then $\frac{T_t^2}{k_t}$ is constant. This provides the condition to be maintained to keep the average heat dissipation zero in quasistatic regime along the adiabatic steps, which we define as microadiabaticity here. This condition of microadiabaticity has been theoretically derived and experimentally verified in \cite{martinez,ignacio}. }}

\section{A Microadiabatic Protocol}

{\textcolor{black}{In this section we will construct a class of micro adiabatic protocol involving $k_t$ and $T_t$. A particular protocol of this class is theoretically as well as experimentally realised in \cite{martinez,ignacio}. While deriving the protocol we will closely follow \cite{martinez,ignacio}.  We consider that the first step of the protocol to be isothermal compression with the bath of temperature $T_c$ within time $0<t<\frac{\tau}{4}$ where $k(0)=k_0$ and $k(\tau/4)=k_1$ with $k_1>k_0$. Next step is an adiabatic compression where the temperature $T_c$ goes to $T_h$ and $k_1$ is raised to $k_2$ within $\frac{\tau}{4}<t<\frac{\tau}{2}$. Third step is an isothermal expansion at $T=T_h$ and $k_2$ decreases to $k_3$ when $\frac{\tau}{2}<t<{\tau^*}$ where $\tau^*$ will be fixed  by known parameters. The fourth and final step is an adiabatic expansion where the temperature $T_h$ decreases to $T_c$ and $k_3$ decreases to $k_0$ within ${\tau^*}<t<{\tau}$. From the condition of microadiabaticity, $\frac{T_h}{T_c}=\sqrt{\frac{k_2}{k_1}}=\sqrt{\frac{k_3}{k_0}}$. We assume $k(t)=\alpha t^n+k_0$ when $t<\frac{\tau}{2}$ and $k(t)=\alpha (\tau-t)^n+k_0$ when $t >\frac{\tau}{2}$. Here $n$ can be any positive number. Maintaining the continuity
of the protocol we found $\alpha=(\frac{2}{\tau})^n(k_2-k_0)$. Next, together with microadiabaticty, we find $k_1=\frac{k_2+(2^n-1)k_0}{2^n}$}, $k_3=\frac{2^nk_2k_0}{k_2+(2^n-1)k_0}$ and $\tau^*=\tau\left[1-\frac{1}{2}\left(\frac{(2^n-1)k_0}{k_2+(2^n-1)k_0}\right)^{1/n}\right]$. Therefore for given $k_0$, $k_2$ and $\tau$ the microadiabatic protocol $k(t)$ is fixed with the following time dependence of the themperature : $T=T_c$ for $t\leq\frac{\tau}{4}$, $T=T_c\sqrt{\frac{2^n(\alpha t^n+k_0)}{k_2+(2^n-1)k_0}}$ for $\frac{\tau}{4}\leq t\leq\frac{\tau}{2}$, $T=T_h$ for $\frac{\tau}{2}\leq t\leq {\tau}^*$ and finally $T=T_c\sqrt{\frac{\alpha(\tau-t)^n+k_0}{k_0}}$ for ${\tau}^*\leq t\leq \tau$. Note that, for $\tau\leq\frac{\gamma}{m}$ as the
system will not be in quasistatic regime, particle will not be in canonical state and therefore the above protocol will not be micro adiabatic anymore.}   


\section{Results} 

{\textcolor{black}{We solve the equation of motion of the particle with protocol having $n=2$ for $x$ and $v$ with velocity-Verlet algorithm using time-axis discretisation$\Delta t=10^{-3}$. From the equation of motion we obtain first law of stochastic thermodynamics as $\Delta u=w-q$ where $\Delta u=\frac{1}{2}(mv^2+k_tx^2)$ is the energy of the particle at time $t$, $w=\int\frac{\partial u}{\partial t}dt$ is the work done on the particle and $q=\int(-\gamma v+\sqrt{\gamma T}\xi)vdt$ is the heat dissipation along a trajectory governed by under-damped stochastic dynamics of the particle. Along the stochastic trajectories, we compute stochastic thermodynamic quantities. We run the dynamics of the particle repeatedly over large number of cycles (N) in time-periodic steady state. The average of the thermodynamic quantities are taken over all the cycles to obtain average of work and heat and internal energy difference ($W, Q, \Delta U$).}}\\ 
{\textcolor{black}{We calculate four contributions (two from isotherms and two from adiabates) of a cycle to the stochastic thermodynamic quantities separately and their averages. Together with $W$ , we obtain $Q_h (Q_c )$ as average heat dissipation along hot (cold) isotherm. We run the stochastic machine by varying ($\tau$ ,$T_h$ ) and
measure $(Q_h , Q_c ,W)$ to obtain the phase diagram in Fig[1, left]. A remarkable difference between micro and macro heat engines emerge from the phase diagram. Unlike the macro-engines, the stochastic machine is not a heat engine for lower cycle times rather it is a heater of type A where $(Q_h, Q_c, W)$ all three are positive, implying the applied work on the system heats up both the reservoirs. Therefore, the heat engines of micro-world are not as reliable as that of in macro-world.}}\\
{\textcolor{black}{Using the trajectory-based definitions of work and heat we define the stochastic efficiency for a single trajectory as $\eta=-\frac{w}{q_h+q_1^{ad}+q_2^{ad}}$ where $q_h$ is the heat dissipated along hot isotherm and $q_1^{ad}, q_2^{ad} $ are heat dissipated along the adiabatic expansion and compression pathways. We compute the distributions of $\eta$ for
two different cycle times (Fig[1], right). The distributions contain values largely deviated from the
mean and the tail can be fitted with $\sim\eta^{-2}$.}}\\
We Average $\eta$ over N to obtain average efficiency $\langle\eta\rangle$ for different $\tau$ in Fig [1, middle]. It shows that for larger $\tau$ as the
system approaches quasistaticity, $\langle\eta\rangle$ approaches towards the Carnot limit. Due to the dominance of largely deviated values in smaller $\tau$ , there is lack of convergence of the data to a single mean. While defining average efficiency as $\bar \eta=-\frac{W}{Q_h}$ , we notice that the variation of $\bar\eta$ with $\tau$ is rather smooth in comparison to $\langle\eta\rangle$ . Though both approaches to Carnot limit with increasing $\tau$ (Fig[1, middle]).

\begin{figure}[ht]
\centering
\includegraphics[width=1.0\columnwidth]{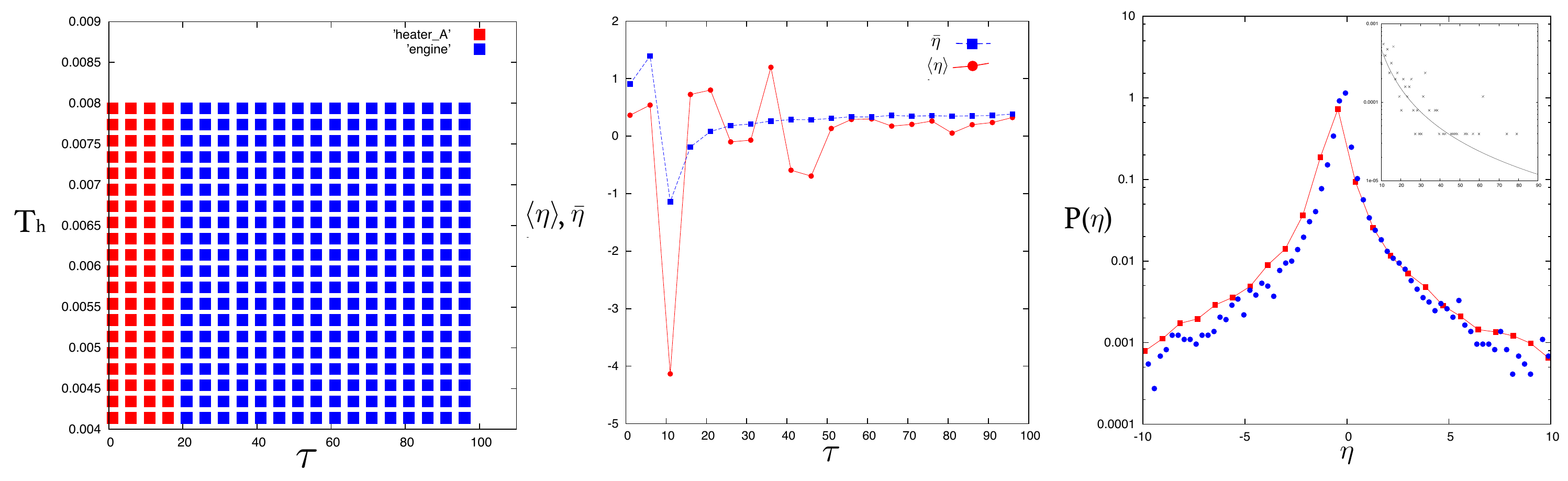}
\caption{Left: Phase diagram. Middle : Average efficiencies with cycle time. Right: Probability distributions efficiencies with cycle time 40 (red) and 60 (blue). In the inset the efficiency distribution at $\tau=40$ is best-fitted with $\sim \eta^{-1.98}$.}
\end{figure}

\section{Conclusion}

We have studied the thermodynamics of single Brownian particle heat engine under microadiabaticity via (i) exploring its different modes of operations by varying cycle times and hot bath temperature (ii) behaviour of averaged efficiencies with varying cycle time and (iii) stochastic efficiency distributions with two different cycle times. For small cycle times the machine will not operate as a heater. With larger cycle time, average stochastic efficiencies will converge to the Carnot limit. The distribution of stochastic efficiency has a power law decay with power close to  (-2) of $\eta$ .This findings signify the role fluctuations play in the small scale. As the the protocol is experimentally realised in \cite{martinez}, our results can be verified experimentally.

\section{Acknowledgement} 


AS thanks UGCFRP and RMS thanks DST, India for financial support. AMJ also thanks DST, India for J. C. Bose National Fellowship. AS thanks Edgar Roldan and I. A. Martinez for Initial discussions.

\end{document}